\documentstyle[preprint,aps,epsf]{revtex}
\tighten
\begin{document}
\draft
\title{Diagrams for Heat Kernel Expansions.}
\author{Ian G. Moss and Wade Naylor}
\address{
Department of Physics, University of Newcastle Upon Tyne, NE1 7RU U.K.
}
\date{March 1999}
\maketitle
\begin{abstract}
A diagramatic heat kernel expansion technique is presented. The method
is especially well suited to the small derivative expansion of the heat
kernel, but it can also be used to reproduce the results obtained by
the approach known as covariant perturbation theory. The new technique
gives an expansion for the heat kernel at coincident points. It can
also be used to obtain the derivative of the heat kernel, and this is
useful for evaluating the expectation values of the stress-energy
tensor.
\end{abstract}
\narrowtext
\section{INTRODUCTION}

The heat kernel was first introduced into quantum field theory to
help evaluate the polarization of the quantum vacuum caused by an
external field \cite{schwinger}. The heat kernel method invented by
Schwinger was developed by DeWitt into one of the key 
techniques for evaluating the one loop effective action in the 
background field approach \cite{dewitt,fulling}.

In many applications, the heat kernel is replaced by its asymptotic 
series in powers of the proper time
\cite{dewitt,seeley,gilkey1}. The coefficients in this series relate
directly to physical quantities, with the first coefficient 
corresponding to the Weyl formula for the spectral density and 
further coefficients being relevant to the renormalisation 
group and to anomalies \cite{deser}. The coefficients depend on 
external potentials, the field strengths of external gauge fields 
and the Riemannian curvature of the underlying space. The first 
six coefficients have been calculated by a wide variety of methods \cite{dewitt,gilkey,avramidi,van,fliegner}.

In practice, the proper-time expansion of the heat kernel is usually
inadequate for obtaining a good approximation to the effective action
but an expansion in powers of
covariant derivatives of the background fields may suffice. The
terms in such a series can be organised in various ways depending on
which quantities are to be regarded as being small. If the field 
strengths and derivatives of the external potentials are small then an
expansion in powers of derivatives will be appropriate. This expansion 
is essentially a low energy expansion.

Schwinger's original calculation of the vacuum polarization in a
constant external field corresponds to a slightly different type of
expansion where derivatives of the field strength can be neglected but
the field strength itself may be large. This type of expansion leads
directly to the Euler-Heisenberg effective action.

A complimentary situation would be where the field strengths
where small compared to their derivatives. The expansion should then be
arranged in increasing powers of the curvatures with infinite numbers of derivatives resummed at each stage. This large derivative or high 
energy expansion is possible and it is used in an approach known 
as covariant perturbation theory \cite{vil,bar}.

In this paper we present a new diagramatic expansion technique, which
is basically a small derivative expansion of the heat kernel, but which
can also be used to reproduce the results obtained by covariant 
perturbation theory. The new technique gives an
expansion for the heat kernel at coincident points. It can also be used
to obtain local derivatives of the heat kernel, and this is useful for
evaluating the expectation values of the stress-energy tensor
\cite{hawking}.

There are close affinities between our method and the method invented
by Barvinski and Vilkoviski \cite{bar} and developed by Barvinski,
Osborn and Gusev \cite{osborn}. In all of these methods, the diagrams
are essentially the Feynman diagrams for a non-relativistic particle
interacting with a background potential. However, there are important
differences between the method presented here and previous methods for
expanding the heat kernel. Firstly, we have organised our expansion
into powers of derivatives of the potential, rather than the proper 
time. Our rules give the coincident heat kernel, wereas results in 
\cite{bar} are given only for the integrated heat kernel. Furthermore, 
our rules apply in curved space, whereas results in \cite{osborn} are 
for flat space, although a curved space generalisation was sketched out.

We will present various new results including new terms in the
derivative expansion in flat space and for constant background gauge
field strengths. We also give new terms in the normal coordinate
expansion of the metric that can be used elsewhere.

\section{THE BASIC METHOD}

Consider a Riemannian manifold $({\cal M},g)$, and a second order
operator $\Delta$ acting on a set of fields $\phi(x)$. We are
interested in the heat kernel $G(x,x',t)$, defined by
\begin{equation}
\Delta G+\dot G=0,\qquad G(x,x',0)=\delta(x,x')\label{heat}
\end{equation}
where a dot denotes derivatives with respect to the proper time $t$ and
$\delta(x,x')$ is the covariant delta function.

Throughout this section the operator will be taken to have a specific
form
\begin{equation}
-g^{\mu\nu}\partial_\mu\partial_\nu+\Gamma^\mu\partial_\mu+W\label{op}
\end{equation}
where $\partial_\mu$ is the partial derivative with respect to $x^\mu$.
The aim of this section is to obtain a series expansion for the heat
kernel in powers of partial derivatives of $g^{\mu\nu}$, $\Gamma^\mu$
and $W$.

Our method begins in the way introduced by Parker \cite{parker} and
follows some steps of reference \cite{moss}. We define
\begin{equation}
K(k,x',t)=\int d\mu(x)\ e^{-ik(x-x')}G(x,x',t)
\end{equation}
Taking the Fourier transform of equation (\ref{heat}) and integrating
by parts gives,
\begin{equation}
\dot K(k,x',t)=-\int d\mu(x) (k^2+W-Z)\,G(x,x',t)\,e^{-ik(x-x')}
\end{equation}
\begin{equation}
Z(k,x,x')=(\partial_\mu-ik_\mu)(\partial_\nu-ik_\nu)h^{\mu\nu}
+(\partial_\mu-ik_\mu)\Gamma^\mu-W(x)+W(x')\label{zed}
\end{equation}
and $h^{\mu\nu}=g^{\mu\nu}-\delta^{\mu\nu}$. Expanding $Z$ in a Taylor
series about $x=x'$ gives
\begin{equation}
\dot K(k,x',t)=-\left(k^2+W-Z(k,D)\right)\,K(k,x',t) \label{hot}
\end{equation}
where
\begin{equation}
Z(k,D)=\sum_{r=0}^\infty {1\over r!}Z_{,\mu_1\dots\mu_r}
D^{\mu_1}\dots D^{\mu_r}
\end{equation}
and $D^\mu=i\partial/\partial k_\mu$.

Equation (\ref{hot}) has a solution in terms of a time-ordered
exponential,
\begin{equation}
K=e^{-(k^2+W)t}\,T\exp\left(\int_0^t e^{(k^2+W)t'}
Z(k,D)e^{-(k^2+W)t'}dt'\right)
\end{equation}
Redistributing some of the terms gives a more convenient form
\begin{equation}
K=e^{-Wt}e^{-k^2t/2}\,T\exp\left(\int_0^t e^{Wt'}
Z(\case1/2i\delta',\delta)e^{-Wt'}dt'\right)e^{-k^2t/2}\label{redhot}
\end{equation}
where
\begin{equation}
\delta^\mu(t')=D^\mu-2ik^\mu t'+ik^\mu t\label{delt}
\end{equation}
and $\delta'=-2ik$. The function $Z$ can be simplified by using the
identity
\begin{equation}
 \lbrack
 F(\delta),\delta'_\mu\rbrack=2F(\delta)_{,\mu},\label{identity}
\end{equation}
which is valid for any function $F(\delta)$, to remove some of the derivatives.
After this simplification,
\begin{equation}
Z=\case1/4h^{\mu\nu}(\delta)\delta'_\mu\delta'_\nu+
\case1/2\Gamma^\mu(\delta)\delta'_\mu
-W(\delta)\label{genv}
\end{equation}
The coincidence limit of the heat kernel is obtained by integration
\begin{equation}
G(x,x,t)=\int d\mu(k) K(k,x,t)\label{measure}
\end{equation}
where $d\mu(k)=d^nk/(2\pi)^n$ in $n$ dimensions. Coincidence limits of
derivatives of the heat kernel can be obtained in a similar way, for
example
\begin{eqnarray}
\left[\partial_\mu G\right]&=&-\int d\mu(k) \case1/2\delta'_\mu
K(k,x,t)\\
\left[\partial_{\mu'} G\right]&=&\int d\mu(k)
K(k,x,t)\case1/2\delta'_\mu
\label{mdiv}
\end{eqnarray}
where $[\dots]$ denotes evaluation at $x'=x$. (The operators in
equation (\ref{redhot}) act on the $\delta'$ in equation (\ref{mdiv})).
Extra derivatives introduce extra factors of $\delta'$.

We shall define
\begin{equation}
\langle \dots\rangle={1\over K_0(t)}\int d\mu(k)e^{-k^2t/2}\dots
e^{-k^2t/2}
\end{equation}
where the normalisation factor
\begin{equation}
K_0(t)=(4\pi t)^{-d/2}\label{kzero}
\end{equation}
for $d$ dimensions. The coincident limit of the heat kernel is
therefore
\begin{equation}
G(x,x,t)=K_0(t)e^{-Wt}\left\langle T\exp\left(\int_0^t e^{Wt'}
Z(\case1/2i\delta',\delta)e^{-Wt'}dt'\right)\right\rangle.\label{heater}
\end{equation}
Equation (\ref{heater}) is the basic equation for obtaining the
derivative expansion of the heat kernel. It yields a convenient
diagramatic expansion by the conventional route using Wicks theorem. It
is similar to an equation (for a more restriced operator) used in 
reference \cite{osborn}.

We can expand the exponential and introduce creation and anihilation
operators $c_\pm(k,D)$ that satisfy
\begin{equation}
c^\mu_-\,e^{-k^2t/2}=0,\qquad [c^\mu_-,c^\nu_+]=\delta^{\mu\nu}
\end{equation}
The operators $\delta$ and $\delta'$ can then be replaced by $c_\pm$,
\begin{equation}
\delta^\mu(t')=\sqrt{2\over t}\{(t-t')c_-^\mu+t'c^\mu_+\}\label{delta}
\end{equation}
%
%
By Wick's theorem, the time-ordered product of operators can be
replaced by products of propagators. There are three different
propagators corresponding to the distinct ways of combining the 
operators,
\begin{eqnarray}
\langle T\delta^\mu(t_i)\delta^\nu(t_j)\rangle&=&
\delta^{\mu\nu}D(t_i,t_j)\\
\langle T\delta'_\mu(t_i)\delta^\nu(t_j)\rangle&=&
\delta_\mu{}^\nu\overrightarrow D(t_i,t_j)\\
\langle T\delta'_\mu(t_i)\delta'_\nu(t_j)\rangle&=&
\delta_{\mu\nu}\stackrel{\leftrightarrow}{D}(t_i,t_j)
\end{eqnarray}
These can be evaluated using the creation and anihilation operators,
\begin{eqnarray}
D(t_i,t_j)&=&
2\hbox{min}(t_i,t_j)-2t^{-1}t_it_j\\
\overrightarrow D(t_i,t_j)&=&
2\theta(t_j-t_i)-2t^{-1}t_j\\
\stackrel{\leftrightarrow}{D}(t_i,t_j)&=&
-2t^{-1}
\end{eqnarray}
where $\theta(t)=1$ for $t\ge 0$ and zero otherwise.

After applying Wick's theorem, a typical term in the expansion of
equation ({\ref{heater}) will take the form
\begin{equation}
K_0(t)e^{-Xt}\int dt_1\dots dt_n T\left(
e^{Xt_1}Z_1e^{-Xt_1}\dots e^{Xt_n}Z_ne^{-Xt_n}\right)g(t_1\dots t_n)
\end{equation}
where $g(t_1\dots t_n)$ is a combination of propagators and index
contractions. The factors $Z_i$ arise from the Taylor series expansion
of $Z$ (see (\ref{genv})). When defining these factors it is convenient
to multiply each term in the Taylor series by $p!r!$, where $p$ is the
order in $\delta'$ and $r$ the order in $\delta$, as this simplifies
the combinatoric factors in the final expansion.

The terms in the heat kernel expansion can be represented by diagrams,
with lines representing propagators and vertices representing the terms
$Z_i$. There are three types of vertex, corresponding to the three
terms in equation (\ref{genv}). Each term is divided by a numerical factor
corresponding to the order of the symmetry group of the diagram in the
usual way.

\section{COVARIANT EXPANSIONS}

Second order operators are often written in the form
\begin{equation}
\Delta=-D^2+X\label{covop}
\end{equation}
where $D=\nabla+iA$ is a gauge covariant derivative, with gauge
conection $A$ and Levi-Civita connection $\Gamma$. The heat kernel
expansion can then be expressed in terms of curvatures and covariant
derivatives. We will use the notation $O(D^n)$ to describe the size of
any term in the heat kernel expansion which has $n$ covariant
derivatives and any number of factors of $X$. The field strengths will
be formally $O(D)$.

The partial derivative expansion can be obtained from the method just
described. With this particular operator the vertices are generated by
a function $\bar Z$, where (see equation (\ref{genv})),
\begin{eqnarray}
\bar Z&=&\case1/4h^{\mu\nu}\delta'_\mu \delta'_\nu+
\case1/2\Gamma^\mu\delta'_\mu-
\case1/2i\delta^{\mu\nu}\{\delta'_\mu,A_\nu\}-X\nonumber\\
&&-\case1/2i\{h^{\mu\nu},\delta'_\mu\}A_\nu
-i\Gamma^\mu A_\mu-g^{\mu\nu}A_\mu A_\nu\label{covz}
\end{eqnarray}
We have made use of the identity (\ref{identity}) in this expression.
The heat kernel expansion is generated by expanding equation
(\ref{heater}). The anticommutators $\{,\}$ only affect contractions at
equal times, and this is taken into account in the diagrammatic rules
given below.

In principle, each of the terms generates a series of vertices when
expanded in powers of $\delta$. However, most of the terms in $\bar Z$
can be eliminated if the propagator $\stackrel{\leftrightarrow}D$
attached to a vertex formed from the gauge field $A$ is replaced by the
propagator
\begin{equation}
\stackrel{\leftrightarrow}{D}_m(t_i,t_j)=2\delta(t_i-t_j)-2t^{-1}
\end{equation}
The vertices are generated by
\begin{equation}
Z=\case1/4 h^{\mu\nu}\delta'_\mu\delta'_\nu+
\case1/2\Gamma^\mu\delta'_\mu-
\case1/2i\delta^{\mu\nu}\{\delta'_\mu,A_\nu\}-X\label{zg}
\end{equation}
This is particularly useful because the new propagator is the
derivative of $\overrightarrow D$. The details are given in appendix
A.

Each of the vertices in the diagramatic expansion corresponds to a
multiple partial derivative of $Z$. It is convenient to define vertex
functions by
\begin{equation}
D_\xi^rZ=\xi^{\mu_1}\dots\xi^{\mu_r} Z_{,\mu_1\dots\mu_r}
\end{equation}
In a system of normal coordinates the vertex functions can also be
expressed in terms curvatures and field strengths, as desribed in
appendix B.

The rules for evaluating the heat kernel $G(x,x,t)$ consist of drawing
all possible diagrams with lines from figure \ref{fig1} and vertices
from figure \ref{fig2}. The $i$'th vertex is associated with its own
time variable $t_i$, where $0\le t_i\le t$. The expression
corresponding to the diagram has

(1) For each line, stretching from a vertex $i$ to vertex $j$, one of
the propagators
\begin{eqnarray}
D(t_i,t_j)&=&2\hbox{min}(t_i,t_j)-2t^{-1}t_it_j\label{propa}\\
\overrightarrow D(t_i,t_j)&=&
2\theta(t_j-t_i)-2t^{-1}t_j\\
\stackrel{\leftrightarrow}{D}(t_i,t_j)&=&
\beta\delta(t_i-t_j)-2t^{-1}\label{propb}
\end{eqnarray}
depending on the number of arrows. In these expressions, $\theta(0)=1$ 
and $\beta=0$ if only $g$ and $\omega$ vertices are attached, whilst
$\theta(0)=\case1/2$ and $\beta=2$ if there is an $A$ vertex.

(2) For each vertex, the coefficient of the appropriate vertex function
\begin{eqnarray}
&&\case1/2e^{Xt_i}D_\xi^rh^{\mu\nu}e^{-Xt_i}\\
&&\case1/2e^{Xt_i}D_\xi^r\Gamma^\mu e^{-Xt_i}\\
-&&ie^{Xt_i}D_\xi^rA_\mu e^{-Xt_i}\\
-&&e^{Xt_i}D_\xi^rX e^{-Xt_i}
\end{eqnarray}
These vertex functions can be expressed in terms of covariant tensors
by using tables \ref{tab1} to \ref{tab3}. The components of these 
tensors are contracted according to the arrangement of the lines.

The differentiated heat kernel at two coincident points is represented
by diagrams with an external line for each derivative of the heat
kernel. The free ends of the $x$ and $x'$ derivative lines are
associated with the time variables $t$ and $0$ respectively. (This
respects the ordering of the terms in equation (\ref{mdiv})). These
contribute

(3) For each external line, the appropriate propagator from rule (1)
multiplied by $\case1/2$ for an $x'$ and $-\case1/2$ for an $x$ 
derivative.

The resulting expression after applying rules (1)-(3) has to be time
ordered and integrated with respect to the time variables. It is then
multiplied by $K_0\exp(-Xt)$ (see equation (\ref{kzero})) and divided
by the order of the symmetry group of the diagram.

The expansion obtained from these diagrams is a derivative expansion
where a diagram with $n$ internal lines and $m$ arrows produces a term
of $O(D^{2n-m})$.

\section{ABELIAN THEORIES IN FLAT SPACE}

The evaluation of the heat kernel simplifies considerably for abelian
gauge theories in flat space. In some situations it is even possible to
obtain exact results for the heat kernel, for example with a constant
field strength $F$. In this section we will give the leading terms in a
small derivative expansion of the heat kernel and show how resummations
can be performed to generalise Schwinger's result for constant fields.
We also show how resummations can be used to obtain an expansion for
the heat kernel in powers of the curvatures, which can be regarded as a
large derivative or high energy expansion \cite{bar}.

The most significant simplification for abelian theories is that only
the connected diagrams need be evaluated, since the results from the
connected diagrams can be exponentiated to reproduce the results from
the complete set of diagrams. This follows from the same simple
counting argument that applies to the usual Feynman graph expansion.

Other important simplifications include the fact that the vertex
functions can be given explicitly,
\begin{eqnarray}
D_\xi^{n-1}A_\mu&=&{n-1\over n}F_{\nu_1\mu,\nu_2\dots\nu_{n-1}}
\xi^{\nu_1}\dots\xi^{\nu_{n-1}}\\
D_\xi^nX&=&X_{,\nu_1\dots\nu_n}\xi^{\nu_1}\dots\xi^{\nu_n}
\end{eqnarray}
We can also reduce the number of diagrams by integration by parts. This
is possible because of the relationships
\begin{eqnarray}
\overrightarrow D(t_i,t_j)&=&{\partial\over\partial t_i} D(t_i,t_j)\\
\stackrel{\leftrightarrow}{D}(t_i,t_j)&=&{\partial\over\partial t_j}
\overrightarrow D(t_i,t_j)\label{prel}
\end{eqnarray}
that enable us to eliminate diagrams with
$\stackrel{\leftrightarrow}{D}$.

The diagrams for some terms in the small derivative expansion with
$F=0$ are shown in figure \ref{fig3}. We write the small derivative
expansion of the heat kernel in the form
\begin{equation}
G(x,x,t)=K_0(t)\exp\left(\sum_{n=0}^\infty W_n\right)
\end{equation}
Results up to $O(D^6)$ can be obtained by drawing all diagrams with up
to three lines and using table \ref{tab5} to evaluate the time
integrals,
\begin{eqnarray}
W_0&=&-Xt\\
W_2&=&-\case1/6(\partial^2X)t^2+\case1/{12}(\partial X)^2t^3\\
W_4&=&-\case1/{60}(\partial^4X)t^3+
\case1/{90}(\partial_\mu\partial_\nu X)
(\partial^\mu\partial^\nu X)t^4\nonumber\\
&&+\case1/{30}(\partial^\mu X\partial_\mu\partial^2 X)t^4
-\case1/{30}(\partial^\mu X)(\partial^\nu X)
(\partial_\mu\partial_\nu X)t^5
\\
W_6&=&-\case1/{840}(\partial^6X)t^4
+\case1/{480}(\partial_\mu\partial^4 X)(\partial^\mu X)t^5\nonumber\\
&&+\case1/{840}(\partial_\mu\partial_\nu\partial_\rho X)
(\partial^\mu\partial^\nu\partial^\rho X)t^5
+\case1/{210}(\partial_\mu\partial_\nu \partial^2X)
(\partial^\mu\partial^\nu X)t^5\nonumber\\
&&+\case{17}/{5040}(\partial_\mu\partial^2 X)
(\partial^\mu\partial^2 X)t^6
-\case{17}/{2520}(\partial_\mu X)(\partial_\nu\partial^2 X)
(\partial^\mu\partial^\nu X)t^6\nonumber\\
&&-\case1/{480}(\partial_\mu X)(\partial_\nu X)
(\partial^\mu\partial^\nu \partial^2X)t^6
-\case1/{210}(\partial_\mu X)(\partial_\nu\partial_\rho X)
(\partial^\mu\partial^\nu\partial^\rho X)t^6\nonumber\\
&&-\case8/{2835}(\partial^\mu\partial_\nu X)
(\partial^\nu\partial_\rho X)
(\partial^\rho\partial_\mu X)t^6
+\case1/{840}(\partial_\mu X)(\partial_\nu X)(\partial_\rho X)
(\partial^\mu\partial^\nu\partial^\rho X)t^7\nonumber\\
&&-\case{17}/{5040}(\partial_\mu X)(\partial_\nu X)
(\partial^\mu\partial^\rho X)
(\partial^\nu\partial_\rho X)t^7
\end{eqnarray}
This exponentiated form also correctly reproduces any terms that can be
reduced to products of $O(D^6)$ terms. As might be expected, most of
the terms in $W_4$ are also recovered in the proper time expansion of
the heat kernel \cite{avramidi}, or by other means \cite{osborn}.
However, most terms in $W_6$ are new and were obtained with very little
effort.

The fact that the field strength has been regarded as a small quantity
might be regarded as a drawback in the present approach. This can be
overcome by a resummation of field strength terms to produce a small
derivative expansion where $F$ is of order $D^0$.

First of all define a new propagator $D_F$,
\begin{equation}
D_F(t_r,t_s)=\sum_{n=0}^\infty D_n(t_r,t_s)F^n
\end{equation}
where $F$ is a matrix and
\begin{equation}
D_n(t_r,t_s)=\int dt_1\dots dt_n D(t_r,t_1)\overrightarrow D(t_1,t_2)
\dots\overrightarrow D(t_n,t_s)
\end{equation}
We also define $\overrightarrow D_F$ by differentiating with respect to
$t_r$. An explicit expression for $D_F$ and is given in appendix C.

The contribution to the heat kernel from ring diagrams involving only
the field strength is
\begin{equation}
W_0=\sum_{n=1}^\infty{1\over 2n}{\rm tr}(F^n)
\int_0^t\overrightarrow D_{n-1}(t',t')dt'
\end{equation}
We can write this in terms of the new propagator,
\begin{equation}
W_0={1\over 2}\int_0^Fd\omega\int_0^t dt'\overrightarrow
D_\omega(t',t')
\end{equation}
By equation (\ref{green}),
\begin{equation}
W_0=-{1\over 2}{\rm tr}\log{\sinh Ft\over Ft}
\end{equation}
This recovers Schwinger's result for the heat kernel in a background
with constant field strength \cite{schwinger}.

In other diagrams, replacing the propagator $D$ by $D_F$ resums all of
the terms involving $F$. For example,
\begin{equation}
W_2=\case1/2\int dt_1 dt_2X_{,\mu}X_{,\nu} D_F(t_1,t_2)^{\mu\nu}
-\case1/2\int dt_1 {\rm tr}X_{,\mu\nu}D_F(t_1,t_1)^{\mu\nu}.
\end{equation}
Again, after using equation (\ref{green}),
\begin{equation}
W_2=\left(\case1/4X_{,\mu}X_{,\nu}-\case1/2X_{,\mu\nu}\right)
\left(F^{-1}t^2\coth Ft-F^{-2}t\right)^{\mu\nu}
\end{equation}
This new result reduces to the previous result if $F=0$.

A quite different situation exists when the derivatives are large,
$\partial^2X\gg X^2$. The large derivative expansion of the heat kernel
is then given by the diagrams shown in figure \ref{fig4}. By summing
these diagrams it is possible to order the derivative expansion in
powers of $X$, thus
\begin{equation}
G(x,x,t)\sim G_1(x,x,t)+G_2(x,x,t)+\dots
\end{equation}
where
\begin{eqnarray}
G_1(x,x,t)&=&K_0(t)\int d\mu(k_1)(-1)e^{ik_1\cdot x}
f(ik_1t)t\hat X(k_1)\label{gone}\\
G_2(x,x,t)&=&K_0(t)\int d\mu(k_1)d\mu(k_2) e^{i(k_1+k_2)\cdot x}
f(ik_1t,ik_2t)t^2\hat X(k_1)\hat X(k_2)\label{gtwo}
\end{eqnarray}
We have written $\hat X(k)$ for the Fourrier transform of $X$.
According the the diagrammatic rules, the single vertex diagrams give
\begin{equation}
f(ik_1t)=\sum_p{1\over 2^p}{1\over p!}\int {dt_1\over
t}D(t_1,t_1)^p(ik_1)^{2p}
\end{equation}
Similarly, for the two-vertex diagrams
\begin{eqnarray}
&f(ik_1t,ik_2t)&=\\\nonumber
&&\sum_{p,q,r}{1\over2^{q+r}}{1\over q!r!p!}
\int {dt_1\over t}{dt_2\over t} D(t_1,t_1)^qD(t_1,t_2)^pD(t_2,t_2)^r
(ik_1)^{2q}(ik_2)^{2r}(-k_1\cdot k_2)^p
\end{eqnarray}
Using the explicit form of the propagator (\ref{propa}), we can perform 
the sumation to express $f(a)$ and $f(a,b)$ in closed form,
\begin{eqnarray}
f(a)&=&\int_0^1 dx\,e^{x(1-x)a^2}\\
f(a,b)&=&2\int_0^1dx_1\int_0^{1-x_1}dx_2\, e^{x_1(1-x_1)a^2
+x_2(1-x_2)b^2+2x_1x_2 a\cdot b}.
\end{eqnarray}
The local expression for the heat kernel in the large derivative limit
given by equations (\ref{gone}) and (\ref{gtwo}) can be used in any 
region in which the derivatives are large, even if they are not large 
everywhere.

For the integrated heat kernel, $k_1=-k_2$ and we can use $f(a,-a)=f(a)$
to obtain
\begin{equation}
\int d\mu(x) G(x,x,t)=K_0(t)\int d\mu(k)f(ikt)t^2\hat X(k)\hat X(-k)
\end{equation}
This demonstrates agreement between the average of our heat kernel and
the results for the integrated heat kernel given in reference \cite{bar}.

\section{CONCLUSIONS}

We have seen that a diagrammatic approach gives the covariant derivative
expansion of the heat kernel in both the large and small derivative
limits. The method gives the coincident limits of the heat kernel or
derivatives of the heat kernel and can be applied to any minimal second
order operator.

The new diagramatic expansion reproduces known
results quickly and reliably, and we have also obtained some new terms in
the derivative expansion in flat space with constant or vanishing
background gauge field strengths. The derivation of more new results is
presently underway.

One important area of application involves finding the heat kernel at
finite temperatures to evaluate thermodynamic functions. The new method
generalises very easily to this situation by following the ideas
presented in \cite{moss}. If only spatial derivatives are present, then
the only change is the replacement of $K_0(t)$ by a Jacobi theta
function,
\begin{equation}
K_0(t)=(4\pi t)^{(1-d)/2}\theta_3(0,4\pi i\beta^{-2}t)
\end{equation}
where $\beta$ is the inverse temperature.

\appendix

\section{VERTEX ELIMINATION}

In this appendix we examine how some of the vertices in the diagramatic
expansion can be eliminated by modifying the propagator.

The full set of terms which generate the vertices are given by equation
({\ref{covz}),
\begin{equation}\matrix{
Z_g=\case1/4h^{\mu\nu}\delta'_\mu \delta'_\nu,\hfill&
Z_\omega=\case1/2\Gamma^\mu\delta'_\mu,\hfill&
Z_a=-\case1/2i\delta^{\mu\nu}\{\delta'_\mu,A_\nu\},\hfill\cr
Z_{ag}=-\case1/2ih^{\mu\nu}\{\delta'_\mu,A_\nu\},\hfill&
Z_{\omega a}=-i\Gamma^\mu A_\mu,\hfill&\cr
Z_{aa}=-\delta^{\mu\nu}A_\mu A_\nu,\hfill&
Z_{aga}=-h^{\mu\nu}A_\mu A_\nu.\hfill&\cr}
\end{equation}
Modified brackets $\langle\dots\rangle_m$ are defined by applying
Wick's theorem with a modified propagator
\begin{equation}
\stackrel{\leftrightarrow}{D}=2\delta(t_i-t_j)-2t^{-1}
\end{equation}
connecting the $A$ vertices and with a restricted set of vertices
generated by $Z_g$, $Z_\omega$ and $Z_a$ only. Contractions
between the $\delta'$ factors reproduce the missing vertices. Expanding
the exponential in equation (\ref{redhot}) with the reduced set of
vertices gives a typical term
\begin{equation}
{1\over p!q!r!}\langle T\,Z_g^pZ_\omega^qZ_a^r\rangle_m=
\sum_{j\dots m}c_{jklm}\langle T\,
Z_{ag}^jZ_{\omega a}^kZ_{aa}^lZ_{aga}^m
Z_g^{p'}Z_\omega^{q'}Z_a^{r'}\rangle\label{typ}
\end{equation}
where
\begin{eqnarray}
p'&=&p-j-m,\\
q'&=&q-k,\\
r'&=&r-j-k-2l-2m
\end{eqnarray}
and the $c_{jklm}$ are combinatorial factors. These factors can be
found by counting the number of ways in which to divide up the vertices
and then counting the possible contractions. The result is equal to the
coefficient of the identical term obtained by expanding the time
ordered exponential with the full set of vertices and the unmodified
propagator.

\section{NORMAL COORDINATE EXPANSIONS}

It is well known that in a system of normal coordinates the partial
derivatives of the metric and the gauge fields can be replaced by
covariant expressions involving curvatures \cite{parker}. We will give
a brief review here to bring out some important features and to give
the results in a form that can be used in the diagramatic expansion of
the heat kernel.

The gauge covariant derivative along a coordinate basis ${\bf e}_\mu$
will be denoted by $D_\mu$. The curvature operator ${\cal R}$ can be
defined by the covariant derivative acting on a vector field ${\bf
e}_a$,
\begin{equation}
\lbrack D_\mu,D_\nu\rbrack{\bf e}_a={\cal R}({\bf e}_\mu,{\bf
e}_\nu){\bf e}_a
\end{equation}
By allowing the vector ${\bf e}_a$ to point in either the direction 
tangential to the manifold or the direction of the internal symmetry 
allows the connection coefficients to include both the tetrad 
connection and the gauge field $A_\mu$. The corresponding components 
of the curvature operator are the Riemann tensor and the Field 
strength tensor.

We choose a point $P$ and an orthonormal frame ${\bf e}_a$ at $P$ to
set up the normal coordinates. Consider a family of geodesics passing
through $P$ with tangent vectors $\xi$. The normal coordinates $x^\mu$
of a point $Q$ are defined by
\begin{equation}
x^\mu=\sigma\xi^\mu(\tau)
\end{equation}
where $\sigma$ is the distance along the geodesic from $P$ to $Q$ and
$\xi(\tau)$ is the unit tangent vector at $P$. This construction
implies that the coordinate vectors ${\bf e}_\mu$ satisfy a 
commutation relation
\begin{equation}
D_\xi(\sigma{\bf e}_\mu)-\sigma D_\mu\xi=\lambda\xi\label{ncom}
\end{equation}
for a constant $\lambda$. We can also arrange that $D_\xi{\bf e}_a=0$
for the reference frame used to calculate the curvature components.

By repeatedly differentiating $\sigma A_\mu$ and $\sigma{\bf e}_\mu$ at
$P$ we get two important identities
\begin{eqnarray}
D_\xi^{n-1}A_\mu&=&{n-1\over n}D_\xi^{n-2}F(\xi,{\bf
e}_\mu)\label{dna}\\
D_\xi^n{\bf e}_\mu&=&{n-1\over n+1}D_\xi^{n-2}R(\xi,{\bf
e}_\mu)\xi\label{dne}
\end{eqnarray}
These can be used recursively to allow us to replace the connection
components and their derivatives with covariant derivatives of the 
field strengths.

If the manifold is flat then equation (\ref{dna}) has an explicit
solution
\begin{equation}
D_\xi^{n-1}A_\mu={n-1\over n}F_{\nu_1\mu;\nu_2\dots\nu_{n-1}}
\xi^{\nu_1}\dots\xi^{\nu_{n-1}}
\end{equation}
where the `;' denotes the gauge covariant derivative. In this case the
choice of gauge is identical to the gauge introduced originally by Fock
and developed by Schwinger \cite{fock,schwinger2}}.

In the curved case, we are interested in derivatives of the inverse 
metric and the connection $\Gamma^\mu$, where
\begin{equation}
\Gamma^\mu=g^{\mu\nu}\omega_{,\nu}-h^{\mu\nu}{}_{,\nu}
\end{equation}
and $\omega=\log|g^{\mu\nu}|^{1/2}$. We can use matrix notation,
writing $e$ for the tetrad and $g$ for the metric, with $g=e^Te$ and
$\omega=-\log|e|$. Covariant expressions for derivatives of the 
determinant and the inverse of the metric at P can then be obtained 
from Faa di Bruno's formula \cite{faa}, for
example
\begin{eqnarray}
D^n\omega&=&\sum_{m=1}^n\sum_{\{a_i\}}{(-1)^m\over m}
{n!\over (1!)^{a_1}\dots(n!)^{a_n}}
\left\{{\rm tr}\left((De)^{a_1}\dots (D^ne)^{a_n}\right)
+{\rm permutations}\right\}\\
D^ng^{-1}&=&\sum_{m=1}^n\sum_{\{a_i\}}(-1)^m
{(m+1)n!\over (1!)^{a_1}\dots(n!)^{a_n}}
\left\{(De)^{a_1}\dots (D^ne)^{a_n}
+{\rm permutations}\right\}
\end{eqnarray}
summed over $a_1+a_2+\dots a_n=m$ and $a_1+2a_2+\dots na_n=n$. 
Expressions for $D^ne$ can be obtained from equation (\ref{dne}). 

Our results are tabulated in tables \ref{tab1} to \ref{tab3}. These can
be partially checked against similar expansions in a paper by van de
Ven \cite{van}, which contains results up to $n=6$. Our ${\bf R}_j$
corresponds to ${j+3\over j+1}{\bf K}_{j+2}$ and ${\bf F}_j$ to
${j+2\over j+1}{\bf Y}_{j+1}$ in his `index-free' notation.

\section{TIME INTEGRALS}

Each diagram ${\cal D}$ in the heat kernel expansion in derivatives
corresponds to an expression of the form
\begin{equation}
G_{\cal D}=K_0(t)e^{-Xt}\int dt_1\dots dt_n T\left(
e^{Xt_1}Z_1e^{-Xt_1}\dots e^{Xt_n}Z_ne^{-Xt_n}\right)g(t_1\dots t_n)
\end{equation}
We will consider the abelian case first.

When the terms commute we have
\begin{equation}
G_{\cal D}=K_0(t)e^{-Xt}Z_1\dots Z_n\int dt_1\dots dt_n\,g(t_1\dots
t_n)
\end{equation}
where $g(t_1\dots t_n)$ is a product of propagators. The integrals can
be reduced recursively by equations such as
\begin{equation}
\int_0^t D(t_1,t_2)^p dt_2={1\over p+1}D(t_1,t_1)^p
\end{equation}
or similar results. We can also replace arrows over the propagators
$D(t_i,t_j)$ by derivatives with respect to $t_i$ for $\rightarrow$ and
$t_j$ for $\leftarrow$. These simplifications reduce the integrals 
for terms up to $O(D^6)$ to those tabulated in table \ref{tab5}.

The results quoted in section 3 can be obtained from the fact that
$D_\omega(t_i,t_j)$ satisfies
\begin{equation}
{\partial^2D_\omega\over \partial t_j^2}
-2\omega{\partial D_\omega\over \partial t_j}=2\delta(t_j-t_i)
\end{equation}
and $D_\omega$ vanishes at $0$ and $t$. Consequently
\begin{equation}
D_\omega(t_i,t_j)=\cases{
\displaystyle
{(e^{2\omega t_j}-1)(e^{-2\omega t_i}-e^{-2\omega t})
\over \omega(1-e^{-2\omega t})}&$t_i>t_j$\cr
\displaystyle
{(e^{2\omega (t_j-t)}-1)(e^{-2\omega t_i}-1)
\over \omega(1-e^{-2\omega t})}&$t_i<t_j$\cr}\label{green}
\end{equation}
Also $\overrightarrow D_\omega(t_i,t_j)=\partial_i D_\omega(t_i,t_j)$.

In the non-abelian case can need to introduce a sum over permuations
$\pi$,
\begin{equation}
G_{\cal D}=K_0(t)e^{-Xt}\sum_\pi F_\pi(L_it)Z_{\pi(1)}\dots Z_{\pi(n)}
\end{equation}
where $L_i$ is the commutator $L_iZ_j=[X,Z_i]\delta_{ij}$ and
\begin{equation}
F_\pi(\alpha_i)=\int_{t_{\pi(1)}>\dots t_{\pi(n)}}dt_1\dots dt_n\,
g(t_1\dots t_n)e^{(\alpha_1t_1+\dots\alpha_nt_n)/t}
\end{equation}
The functions $F_\pi$ depend on the diagram but not on the particular
gauge group.

We can process the results further by using a gauge transformation to
diagonalise the matrix $X$. It is then possible to choose a canonical
basis $E_a$ of the Lie algebra with the property that
\begin{equation}
[X,E_a]=w(a)E_a
\end{equation}
for constants $w(a)$. The vertices which do not commute with $X$ belong
to representations of the Lie algebra of the gauge group,
$Z_i=Z_i^aE_a$, therefore
\begin{equation}
G_{\cal D}=K_0(t)e^{-Xt}\sum_{a_1\dots a_n}
c_{a_1\dots a_n}Z^{a_1}_1\dots Z^{a_n}_n
\end{equation}
where
\begin{equation}
c_{a_1\dots a_n}=\sum_\pi F_\pi(w(a_i)t)E_{a_{\pi(1)}}\dots
E_{a_{\pi(n)}}.
\end{equation}
We see from this expression that the range of values where $F_\pi$
needs to be calculated is restricted by the Lie algebra roots.

\begin{table}
\begin{tabular}{ll}
$n$&$D^ng^{-1}$\\
\hline\hline
2&$-\case2/3 R_0$\\
3&$- R_1$\\
4&$-\case6/5 R_2+\case8/5 R_0^2$\\
5&$-\case4/3 R_3+8 R_0R_1$\\
6&$-\case{10}/7 R_4+\case{100}/7 R_{(0}R_{2)}
+\case{85}/7 R_1^2-\case{160}/{21} R_0^3$\\
7&$-\case3/2 R_5+\case{122}/3 R_{(0}R_{3)}
+\case{357}/5 R_{(1}R_{2)}
+\case{6611}/{45} R_{(0}^2R_{1)}
+\case{661}/9 R_0R_1R_0$\\
8&$-\case{14}/9 R_6+\case{280}/9 R_{(0}R_{4)}
+\case{824}/9 R_{(1}R_{3)}
+\case{308}/5 R_2{}^2
-\case{1596}/5 R_{(0}^2R_{2)}$\\
&$-\case{3052}/9 R_{(0}R_{1)}^2
-\case{868}/9 R_1R_0R_1
-\case{4676}/9 R_0R_2R_0
+\case{628}/5 R_0^4$\\
\end{tabular}
\caption{Expressions for $D_\xi^ng^{-1}$ in terms of curvatures. In
this table, $R_j$ denotes the linear mapping $R_jX=(D_\xi^j{\cal
R})(\xi,X)\xi$, where ${\cal R}$ is the curvature operator. In terms of
components, $D^ng^{\mu\nu}={\bf e}_\mu\cdot(D^ng^{-1}){\bf e}_\nu$ and
${\bf e}_\mu\cdot R_j{\bf
e}_\nu=R_{\mu\nu_1\nu_2\nu;\nu_3\dots\nu_{j+2}}
\xi^{\nu_1}\dots\xi^{\nu_{j+2}}$.}
\label{tab1}
\end{table}

\begin{table}
\begin{tabular}{ll}
$n$&expression\\
\hline\hline
2&$-\case1/3 R_0$\\
3&$-\case1/2 R_1$\\
4&$-\case3/5 R_2+\case2/{15} R_0{}^2$\\
5&$-\case2/3 R_3+\case2/3 R_0R_1$\\
6&$-\case5/7 R_4+\case8/7 R_0R_2
+\case{15}/{14} R_1{}^2
-\case{16}/{63} R_0{}^3$\\
7&$-\case3/4 R_5+4 R_0R_3
+\case9/2 R_1R_2-\case{43}/3 R_0^2R_1$\\
8&$-\case7/9 R_6+\case{20}/9 R_0R_4
+\case{70}/9 R_1R_3+\case{28}/5 R_2{}^2
-\case{272}/{45} R_0{}^2R_2-\case{310}/9 R_0R_1{}^2
+\case{16}/{15} R_0^4$\\
\end{tabular}
\caption{Expressions for $D_\xi^n\omega$, where
$\omega=\log|g^{\mu\nu}|^{1/2}$, are obtained by tracing the entries in
this table.}
\label{tab2}
\end{table}

\begin{table}
\begin{tabular}{ll}
$n$&$D^{n-1}A$\\
\hline\hline
2&$\case1/2F_0$\\
3&$\case2/3F_1$\\
4&$\case3/4F_2+\case1/3F_0R_0$\\
5&$\case4/5F_3+\case4/5F_1R_0+\case2/5F_0R_1$\\
6&$\case5/6F_4+\case5/3F_2R_0+\case5/3F_1R_1
+\case1/2F_0R_2+\case1/6F_0R_0^2$\\
7&$\case6/7F_5+\case{20}/7F_3R_0+\case{30}/7F_2R_1
+\case{18}/7F_1R_2+\case6/7F_1R_0^2+\case4/7F_0R_3
+\case4/{21}F_0R_1R_0+\case2/7F_0R_0R_1$\\
8&$\case7/8F_6+\case{35}/8F_4R_0+\case{35}/4F_3R_1
+\case{63}/8F_2R_2+\case{21}/8F_2R_0^2+\case7/2F_1R_3
+\case7/2F_1R_1R_0$\\
&$+\case7/4F_1R_0R_1+\case5/8F_0R_4
+\case{10}/8F_0R_2R_0+\case{10}/8F_0R_1^2
+\case3/8F_0R_0R_2+\case1/8F_0R_0^3$\\
\end{tabular}
\caption{Expressions for $D_\xi^{n-1}A$ in terms of curvatures and
field strengths. In this table $F_j$ denotes the linear mapping
$F_jX=(D_\xi^j{\cal R})(\xi,X)$ with gauge field strength indices. In
terms of components, $D_\xi^nA_\mu=(D_\xi^nA){\bf e}_\mu$ and $F_j{\bf
e}_\mu=
F_{\nu_1\mu;\nu_2\dots\nu_{j+1}}\xi^{\nu_1}\dots\xi^{\nu_{j+1}}$.}
\label{tab3}
\end{table}

\begin{table}
\begin{tabular}{llll}
vertices&$g$&numerical factor&term\\
\hline\hline
$1$&$D(t_1,t_1)^p$&$2^p(p!)^2/(2p+1)!$&$t^{p+1}$\\
&$D(t_1,t_1)^p\overrightarrow D(t_1,t_1)^{2q}$&
$2^{p}p!(2q)!(p+q)!/(2p+2q+1)!q!$&$t^{p+1}$\\
\hline
$2$&$D(t_1,t_1)D(t_1,t_2)D(t_2,t_2)$&
$17/630$&$t^5$\\
\hline
$n$&$D(t_1,t_2)\dots D(t_n,t_1)$&
$(-2)^{3n-1}B_{2n}/(2n)!$&$t^{2n}$\\
$2n$&$\overrightarrow D(t_1,t_2)\dots \overrightarrow D(t_{2n},t_1)$&
$-2^{2n-1}B_{2n}/(2n)!$&$t^{2n}$\\
\end{tabular}
\caption{The time integrals corresponding to diagrams for an abelian
operator up to $O(D^6)$ can be reduced to the examples tabulated here.
The results are given by the numerical factors multiplied by the terms
in $t$, and $B_n$ are Bernouilly numbers.}
\label{tab5}
\end{table}

\begin{figure}
\begin{center}
\leavevmode
\epsfxsize=20pc
\epsffile{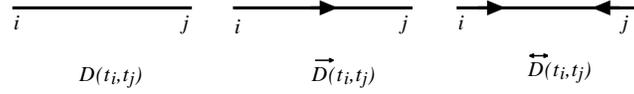}
\end{center}
\caption{Lines appearing in the expansion of the heat equation. They
correspond to the two-point functions of time as indicated.
\label{fig1}}
\end{figure}

\begin{figure}
\begin{center}
\leavevmode
\epsfxsize=30pc
\epsffile{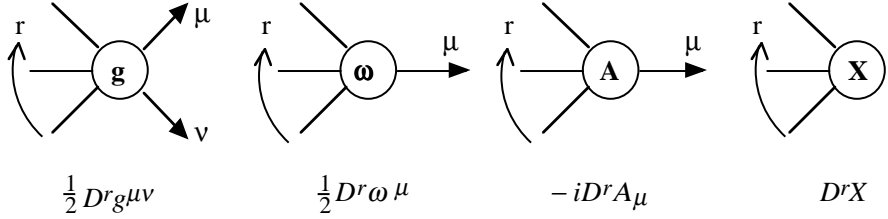}
\end{center}
\caption{Vertices appearing in the diagrammatic expansion. These
vertices are attached to $r$ plain lines and up to 2 lines with arrows.
Each vertex corresponds to an integral over time and the vertex
function indicated below the vertex. Covariant expressions can be found
in the tables. \label{fig2}}
\end{figure}

\begin{figure}
\begin{center}
\leavevmode
\epsfxsize=20pc
\epsffile{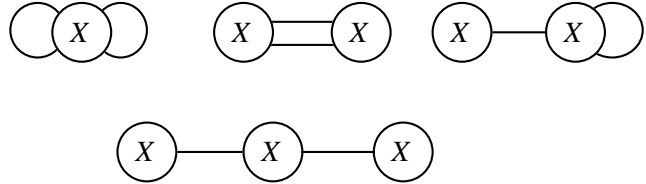}
\end{center}
\caption{The terms in the derivative expansion of $O(D^4)$ are formed
from the diagrams shown here. \label{fig3}}
\end{figure}

\begin{figure}
\begin{center}
\leavevmode
\epsfxsize=20pc
\epsffile{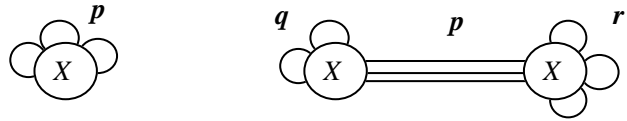}
\end{center}
\caption{Diagrams of this type dominate the large derivative
approximation. \label{fig4}}
\end{figure}


\begin{references}
\bibitem{schwinger}Schwinger J S 1951 {\em Phys. Rev. } {\bf D}82 664;
\bibitem{dewitt}DeWitt B S 1965 {\em Dynamical Theory of Groups and
Fields} (New York: Gordon and Breach)
\bibitem{fulling}Fulling S A (ed) 1995 {\em Discourses in Mathematics
and its Applications} (Heat Kernel Techniques and Quantum Gravity
(Winipeg 1994)) vol 4 (College Station, TX: Department of Mathematics,
Texas A \& M University)
\bibitem{seeley}Seeley R T 1967 {\em Proc. Sympos. Pure Math.} {\bf 10}
{\em Amer. Math. Soc.} 288
\bibitem{gilkey1}Gilkey P B 1984 {\em Invariance Theory, the Heat
Equation and the Atiyah-Singer Index Theorem} (Publish or Perish Inc.,
Wilmington Delaware)
\bibitem{deser}Deser F, Duff M J and Isham C J 1976 {\em Nucl.Phys.}
{\bf B111} 45; Brown L S (1977) {\em Phys. Rev.} {\bf D15} 1469; Dowker
J S and Critchley (1977) {\em Phys. Rev.} {\bf D16} 3390
\bibitem{gilkey}Gilkey P B 1975 {\em J. Diff Geom.} {\bf 10} 601
\bibitem{avramidi}Avramidi I G 1991 {\em Nucl. Phys.} {\bf B355} 712
(erratum 1998 {\em Nucl. Phys.} {\bf B509} 577)
\bibitem{van}van de Ven A E M 1998 {\em Class. Quantum Grav.} {\bf  15}
2311
\bibitem{fliegner}Fliegner D, Haberl P, Schmidt M G and Schubert C 1997
The higher derivative expansion of the effective action by the string
inspired method hep-th/9707189
\bibitem{vil}Vilkoviski G A 1984 {\em in Quantum Theory of Gravity} ed
S M Christensen (Adam Hilger, Bristol 1984)
\bibitem{bar}Barvinski A O and Vilkoviski G A 1987 {\em Nucl. Phys. B}
{\bf 282} 163; Barvinski A O and Vilkoviski G A 1990 {\em Nucl. Phys.
B} {\bf 333} 471
\bibitem{hawking}Hawking S W 1977 {\em Commun. Math. Phys.} {\bf 56}
133
\bibitem{osborn}Barvinski A O, Osborn T A and Gusev Yu V 1995 {\em J.
Math Phys.} {\bf 36} 30
\bibitem{moss}Moss I G and Poletti S (1993) {\em Phys. Rev.} {\bf D47}
5477
\bibitem{feynman}Feynman R P 1948 {\em Rev. Mod. Phys.} {\bf 20} 367
\bibitem{parker}Parker L 1979 {\em Recent Developments in Gravitation
(Cargese 1978))} ed. M Levy and S Deser (New York: Plenum)
\bibitem{fock}Fock V A 1937 {\em Sov. Phys.} {\bf 12} 404
\bibitem{schwinger2}Schwinger J S 1973 {\em Sources, Particles and
Fields} (New York: Addison-Wesley)
\bibitem{faa}de Bruno F 1881 {\em Einleitung in die Theorie der
Bin\"aren Formen} (Leipzig: Verlag B. G. Taubner)
\end{references}
\end{document}